\documentclass[12pt,preprint]{aastex}
\usepackage{color}

\def\chandra{{\em Chandra}\/}
\def\ASCA{{\em ASCA}\/}

\def\XMM{{\em XMM-Newton}\/}

\shorttitle{Weak shock front in A2556}
\shortauthors{Qin et al.}

\begin{document}

\title{ \chandra\ Observation of a  Weak Shock in the Galaxy Cluster A2556}
\author{Zhenzhen Qin\altaffilmark{1}, Haiguang Xu\altaffilmark{1}, Jingying Wang\altaffilmark{1}, Yu Wang\altaffilmark{2}, Junhua Gu\altaffilmark{3} and Xiang-ping Wu\altaffilmark{1,3}}

\altaffiltext{1}{Department of Physics, Shanghai Jiao Tong University, 800 Dongchuan Road, Shanghai 200240, PRC e-mail: qzz@sjtu.edu.cn, hgxu@sjtu.edu.cn}
\altaffiltext{2}{Shanghai Astronomical Observatory, Chinese Academy of Sciences, 80 Nandan Road, Shanghai 200030, PRC}
\altaffiltext{3}{National Astronomical Observatories, Chinese Academy of Sciences, 20A Datun Road, Beijing 100012, PRC}

\begin{abstract}

Based on a 21.5 ks \chandra\ observation of A2556, we identify an edge on the surface brightness profile (SBP) at about 160$h_{71}^{-1}$ kpc northeast of the cluster center, and it corresponds to a shock front whose Mach number $\mathcal{M}$ is calculated to be $1.25_{-0.03}^{+0.02}$. No prominent substructure, such as sub-cluster, is found in either optical or X-ray band that can be associated with the edge, suggesting that the conventional super-sonic motion mechanism may not work in this case. As an alternative solution, we propose that the nonlinear steepening of acoustic wave, which is induced by the turbulence of the ICM at the core of the cluster, can be used to explain the origin of the shock front. Although nonlinear steepening weak shock is expected to occur frequently in clusters, why it is rarely observed still remains a question that requires further investigation, including both deeper X-ray observation and extensive theoretical studies.

\end{abstract}


\section{INTRODUCTION}
Radiative loss via X-ray emission in galaxy clusters has long been supposed to be the origin of the substantial gas inflow, namely the ``cooling flow'' (see Fabian 1994 for a review). However, X-ray spectra from \ASCA\ and \XMM\ observations fail to show line emission from ions with intermediate or low temperatures, which implies that the cooling rate is only at most 20\% of the previously assumed value (e.g., Kaastra et al. 2001; Peterson et al. 2001; Xu et al. 2002; Peterson et al. 2003; Tamura et al. 2003). \chandra\ observations have also confirmed such small cooling rates (McNamara et al. 2000). There must be a heating source in the cluster core region. The most popular candidate of the heating source is the AGN at cluster centers. The AGN dissipates energy via  bubbles, jets, sound waves and weak shocks evolved from sound waves (Fabian \& Sanders 2009). In Perseus and Virgo Clusters, previous works have observed the sound waves and weak shocks generated by activities of central AGNs (Fabian et al. 2003; Ruszkowski et al. 2004; Forman et al. 2005). Besides the AGN activity, some other heating mechanisms have also been concerned in the last decade. Especially since the launch of {\em Chandra}, shock fronts have been revealed clearly in the ICM of galaxy clusters through X-ray imaging-spectroscopic studies, which can efficiently convert kinetic energy of the gas into thermal energy. Such heating process may be associated with the feedback mechanism that balances the radiative cooling in galaxy clusters. 

This work would study the weak shock front in the galaxy cluster A2556 by using the \chandra\ observation. A2556 (RA=23h13m03.3s, Dec=$-$21d37m40s, J2000.0; richness 1) is located at $z \simeq 0.0871$ in the Aquarius supercluster (Batuski et al. 1999), an overdense region of the universe. It is moderately luminous in the X-ray band ($L_{\rm X}\simeq2\times 10^{44}$ erg ${\rm s^{-1}}$; Ebeling et al. 1996), possessing a velocity dispersion $\simeq872$ km ${\rm s^{-1}}$ (White et al. 1997) and a virial mass $\simeq2.5\pm0.1\times10^{15}$ M$_{\odot}$ (Reimers et al. 1996).

We describe the data reduction, imaging and spectroscopy analysis in \S 2; we discuss and summarize our results in \S 3 and  \S 4, respectively. Throughout this paper, we adopt the cosmological parameters $H_{0}=71$ km s$^{-1}$ Mpc$^{-1}$, $\Omega_{m} = 0.27$ and $\Omega_{\Lambda} = 0.73$, so that $1^{\prime}$ corresponds to $96.7h_{71}^{-1}$ kpc at such distance. We utilize the solar abundance standards of Grevesse and Sauval (1998), where the iron abundance relative to hydrogen is $3.16\times10^{-5}$ in number. Unless stated otherwise, all quoted errors are derived at the 68\% confidence level.

\section{DATA ANALYSIS}\label{ana}

\subsection{Observation and Data Reduction}
The \chandra\ observation of A2556 was carried out on October 5, 2001 (ObsID 2226) for a total exposure of 21.5 ks with CCDs 3, 5, 6, 7 and 8 of the Advanced  CCD Imaging Spectrometer (ACIS) in operation. Events were collected  with a frame time of 3.2 s and telemetered in the VeryFaint mode. The focal plane temperature was set to $-120$ $^{\circ}$C. The center of A2556 was positioned nearly on the ACIS-S3 chip with about $0.578^{\prime}$ offset. Most of the X-ray emissions of A2556 are covered by the S3 chip. Therefore, this work focuses on the data drawn from the S3 chip. The \chandra\
data analysis package CIAO software (version 3.4) is used to process the data from the S3 chip. We keep events with \ASCA\ grades 0, 2, 3, 4 and 6, and remove all the bad pixels, bad columns, columns adjacent to bad columns and node boundaries. No background flare is found during the live operating time.

\subsection{X-ray Surface Brightness}\label{xsb}

\begin{figure}
   \begin{center} 
   \includegraphics[angle=0,scale=0.8]{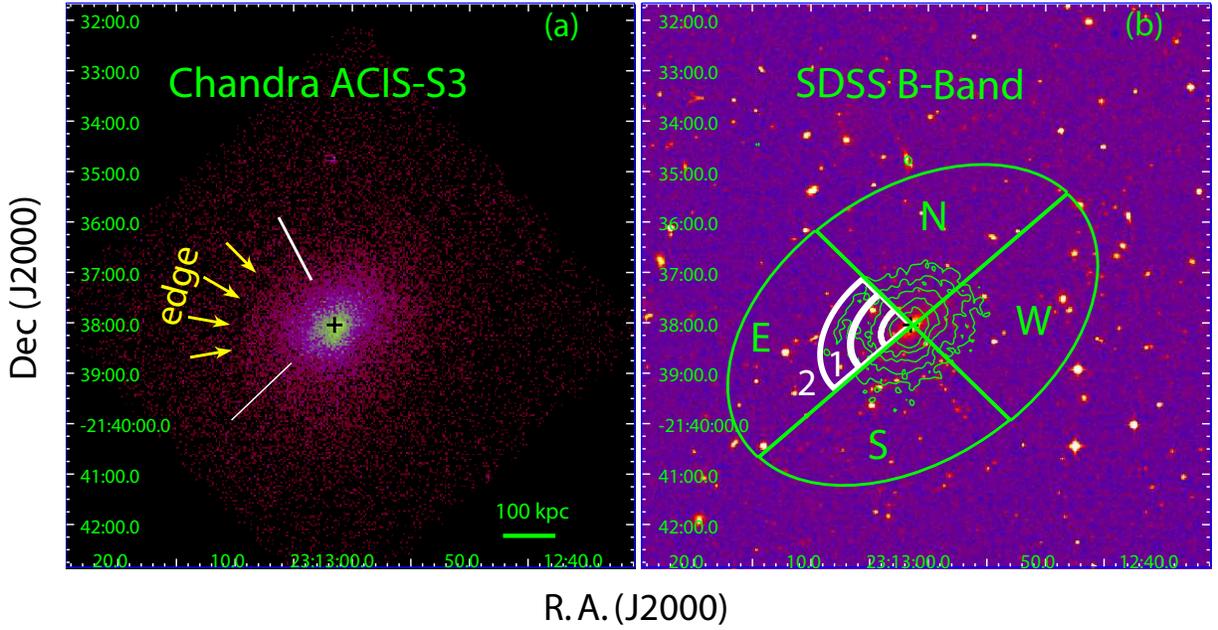}
   \caption{
     (a): Raw \chandra\ image of Abell 2556 in the 0.7--7.0 keV energy band. Two white lines highlight the range of edge span.
     (b): SDSS B-band image for the same sky field, where the X-ray contours are overlaid. The A2556 image is divided into 4 regions noted as E, S, W and N, where SBPs of 4 different directions are extracted. In region E, region 1 and region 2 denote the post-shock and pre-shock regions, respectively. White elliptical annuli in region E demonstrate regions for extracting spectra in \S 2.3.1.
   }\label{ima} 
   \end{center}
   \end{figure}

We show the raw \chandra\  ACIS-S3 image of A2556 in 0.7--7.0 keV band in Figure \ref{ima}a, and have all the point sources excluded, which could be detected at the confidence level of $3\sigma$ by the CIAO tool ${celldetect}$. The black cross in the center of Figure \ref{ima}a indicates the X-ray peak of A2556. The core region of the A2556 shows no complex irregularity and should be composed of a single nucleus; the outer region shows a slight elongation toward the northwest. The X-ray morphology roughly shows a generally relax appearance, except that there exhibits an arm-like structure with an edge at approximately 160$h_{71}^{-1}$ kpc northeast of the cluster center, and the edge spans about $85^{\circ}$ azimuthally from southeast to northeast. We highlight the edge span by two white lines in Figure \ref{ima}a. Figure \ref{ima}b shows the SDSS B-band image of A2556 with the Chandra 0.7--7.0 keV X-ray intensity contours overlaid. The X-ray peak consists with the centroid of the cD galaxy 2MASX J23130142-2138039 within $1^{\prime\prime}$. The cD galaxy shows an elongation in the same direction as the X-ray morphology, which may due to the nearby A2554.

\begin{figure}
    \begin{center} 
     \includegraphics[angle=0, scale=0.45]{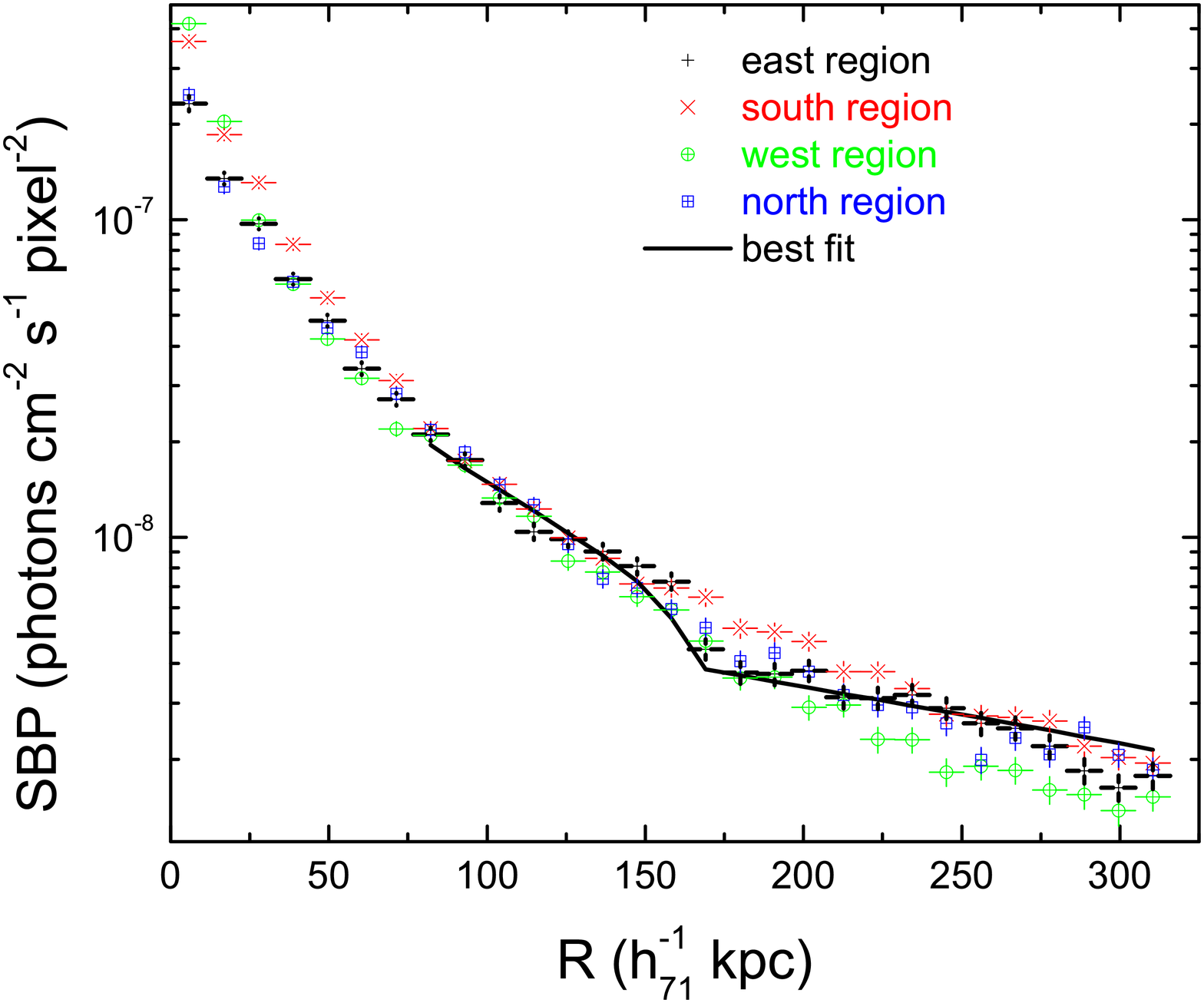}
     \caption{Exposure-corrected SBPs extracted from a series of narrow elliptical annuli across the edge of arm-like structure in region E (see Figure 1b) and other three regions (S, W and N) for comparison. The cross, star cross, circle cross and box cross presents SBPs in regions E, S, W and N, respectively. The solid line demonstrates the best fit of the SBP in region E according to Eq. \ref{er}. A substantial discontinuity of the SBP in region E is located around $R$=160$h_{71}^{-1}$ kpc; in other three regions, no such discontinuity is observed.
       }\label{sbp}
     \end{center}
      \end{figure}

To get more insight into the edge of this arm-like structure, we extract the exposure-corrected  surface brightness profile (SBP) from a series of narrow elliptical annuli, which are parallel to the east (E) region show in the Figure \ref{ima}b. In order to get a comparison, we also extract SBPs of other three directions in Figure \ref{ima}b, namely south (S), west (W) and north (N) regions. The energy band for these profiles is restricted to 0.7--7.0 keV. We show such extracted SBPs in Figure \ref{sbp}. In region E, a substantial discontinuity of the SBP could be observed at approximately 160$h_{71}^{-1}$ kpc, and other three regions do not present such characteristic.

To quantitatively describe the discontinuity of the SBP in region E, we fit observed SBP with a modeled gas emissivity distribution, which corresponds to a function as
\begin{equation}\label{er}
\varepsilon(r) = \left\{ \begin{array}{ll}
 \varepsilon_{1}\left(r/R_{\rm cut}\right)^{\alpha_{1}} & \textrm{$r\le R_{\rm cut}$}\\
 \varepsilon_{2}\left(r/R_{\rm cut}\right)^{\alpha_{2}} & \textrm{$r>R_{\rm cut}$},\\
  \end{array} \right.
\end{equation}
where $r$ represents the radius of the projected SBP; the gas emissivity distribution, $\varepsilon(r)$,  is broken at the truncation radius, $R_{\rm cut}$, and described with two power-law components; $\varepsilon_{1}$, $\varepsilon_{2}$, $\alpha_{1}$ and $\alpha_{2}$ are 4 fitting parameters. The solid line in Figure \ref{sbp} shows the best-fit SBP obtained with Eq. \ref{er}. The best-fit $R_{\rm cut}=158.26^{+0.54}_{-0.42}$$h_{71}^{-1}$ kpc corresponding to the edge of the arm-like structure. $\varepsilon_{1}$ and $\varepsilon_{2}$ shall represent the plasma emissivities inside and outside of the edge, respectively. The best-fit emissivity jump $\varepsilon_{1}/\varepsilon_{2}= 1.90^{+0.02}_{-0.04}$. Since the gas density is related to the emissivity by $\varepsilon \sim \rho^{2}$, the density jump across the edge is calculated to be $\rho_{1}/\rho_{2}=(\varepsilon_{1}/\varepsilon_{2})^{1/2}=1.37^{+0.02}_{-0.03}$.

\subsection{Spectral Analysis}

In the \chandra\ spectral analysis, there are two usual ways to obtain a background spectrum, including cosmic, instrumental and non X-ray particle components. Firstly, the observated dataset can be directly used to extract the background spectrum from a local and uncontaminated region on \chandra\ chips, which must be located far away enough from the source. Secondly, one can draw the background spectrum from the \chandra\ blank field datasets in the same detector region as where the source spectrum is extracted. Since the core radius of A2556 is 350$h_{71}^{-1}$ kpc (White et al. 1997) and it should cover most of ACIS-S3 CCD, we have to utilize the \chandra\ blank-sky template as the background for A2556. The template is tailored to match the actual pointing and the background spectrum is extracted and processed identically to the source spectrum. Then, the background spectrum is rescaled by normalizing its high energy end to the corresponding observed spectrum. Corresponding spectral redistribution matrix files (RMFs) and auxiliary response files (ARFs) are created with CIAO tools, $mkwarf$ and $mkacisrmf$, respectively. All spectra are rebinned to give a minimum of 20 raw counts per spectral bin, which allows $\chi^2$ statistics to be applied. Since the contribution of the hard spectral component is expected to be rather weak, and also to minimize effects of the instrumental background at higher energies and calibration uncertainties at lower energies, the spectrum fitting is restricted to 0.7--7.0 keV .

\subsubsection{Temperature Across the Edge}\label{Tpro}

\begin{figure}
  \begin{center}
   \includegraphics[angle=0,scale=0.23]{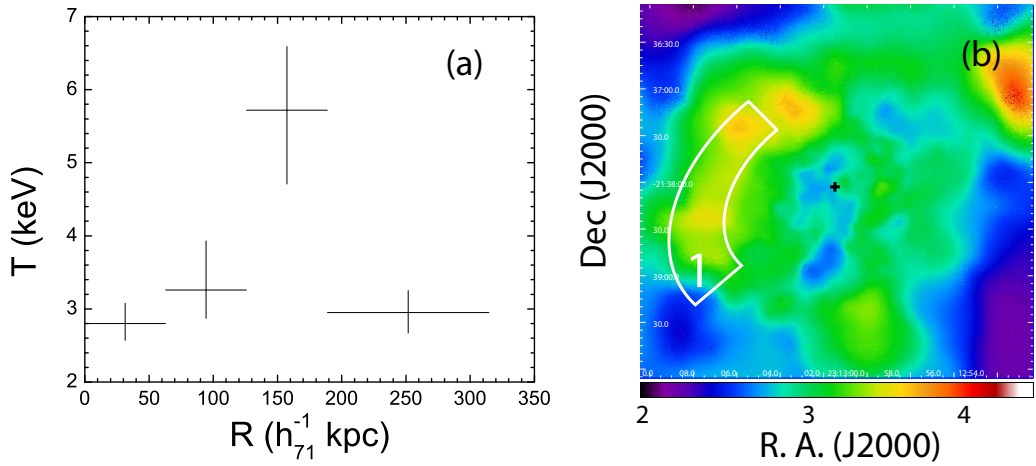}
   \caption{
     (a): Deprojected gas temperature profile across the shock in region E. $R$ represents the distance from the X-ray peak of A2556. A substantial high temperature is observed around $R$=160$h_{71}^{-1}$ kpc, which exactly corresponds to the shock front in A2556.
     (b): The 2-dimensional projected temperature map derived from Gu et al. (2009). The white elliptical annulus corresponds to region 1 in Figure \ref{ima}b.
   }\label{dis_t}
   \end{center}
   \end{figure}

In order to derive the temperature profile across the SBP edge, we extract the spectra from 4 elliptical annuli in region E of Figure \ref{ima}b, so that the SBP edge lies exactly at the boundaries of certain elliptical annulus, and the elliptical annulus covers the azimuthal angle, where the edge is most prominent. We use the XSPEC 12.4.0 package to perform the deprojected spectral analyses. In spectral fitting process, the gas emission is modeled with an absorbed APEC component and the PROJCT model, which performs a 3-dimensional to 2-dimensional projection of prolate ellipsoidal shells onto elliptical annuli and evaluates the influence of outer ellipsoidal shells onto inner ones. The absorption column density $N_{\rm H}$ is fixed at the Galactic value $N_{\rm H}=2.07\times 10^{20}$ cm$^{-2}$ (Dickey \& Lockman 1990). Best-fit temperature values are shown in Figure \ref{dis_t}a. Temperature behind the edge of arm-like structure is much higher than those of ambient regions. The calculated temperature jump across the edge is $T_{1}/T_{2}=1.93^{+0.51}_{-0.72}$. Both gas density and temperature jumps across the edge of the arm-like structure indicate that the edge is a weak shock front.

The Mach number, $\mathcal{M}$, of the weak shock front can be determined by employing the Rankine-Hugoniot shock relation (Landau \& Lifshitz 1959)
\begin{equation} \label{eq:RH1}
  \frac{\rho_{1}}{\rho_{2}} = \frac{(1+\gamma)M^2}{2+(\gamma-1)M^2}
\end{equation}
and
\begin{equation} \label{eq:RH2}
\frac{T_{1}}{T_{2}}=
\frac{\left( 2\gamma M^2 - (\gamma-1) \right) \left( 2+(\gamma-1)M^2\right)}{(\gamma+1)^2 M^2},
\end{equation}
where $T_{1}$ and $T_{2}$ denote the temperature behind (the ``post-shock''; region 1 in Figure \ref{ima}b) and ahead (the ``pre-shock''; region 2 in Figure \ref{ima}b) of the shock, respectively; $\rho_{1}$ and $\rho_{2}$ denote the density behind and ahead of the shock, respectively. For the intracluster gas, $\gamma=5/3$, which is the adiabatic index for monatomic gas. For weak shock fronts, the accuracy of $\mathcal{M}$ derived from the density discontinuity is better than that from temperature discontinuity (Markevitch \& Vikhlinin 2007). Therefore, with the density jump $\rho_{1}/\rho_{2}=1.37_{-0.03}^{+0.02}$ and Eq. \ref{eq:RH1}, we derive the Mach number $\mathcal{M}=1.25_{-0.03}^{+0.02}$. According to Eq. \ref{eq:RH2}, a temperature jump of $T_{1}/T_{2}=1.24_{-0.01}^{+0.02}$ is indicated, which confines the spectroscopic value $1.93^{+0.51}_{-0.72}$ within errors.

We crosscheck the detected temperature jump by examining the 2-dimensional temperature distribution of the gas in A2556. In Figure \ref{dis_t}b, the projected temperature map of A2556 (Gu et al. 2009) exhibits a substantial high temperature region exactly behind the edge of the arm-like structure, which is coincident with the temperature profile shown in Figure \ref{dis_t}a.

\subsubsection{Classification of Weak Cool Core Cluster}\label{cla}

To test whether a cooling flow model fits the spectrum of A2556, we apply four types of models to fit the spectrum in a circle region with radius of $237h_{71}^{-1}$ kpc, which center on the X-ray peak. The circle region covers nearly the whole \chandra\ ACIS-S3 CCD. Firstly, we use the WABS$\times$APEC model (hereafter 1TA) and WABS$\times$MEKAL model (hereafter 1TM) to describe the X-ray emission only by a single isothermal plasma. In these two models, temperatures and elemental abundances are allowed to vary. Secondly, in order to feature a cooling flow, we employ a single temperature MEKAL model plus a multi-phase cooling flow component MCKFLOW, where their abundances are tied together and allowed to vary; the temperature of MEKAL and the higher gas temperature of MCKFLOW are also tied together and allowed to vary. We consider two cases for this model: the lower bound on the gas temperature in the MCKFLOW is fixed at 0.1 keV, which means the cooling material is fully cooled below the X-ray emitting temperature (hereafter denote FC for fully cooling), and this component is set as a free parameter (hereafter denote VC for variable cooling).

Best-fit spectral parameters of four models with 90\% confidence level are presented in Table 1. As a general comment, all models do a reasonably good job at fitting the spectrum with reduced $\chi^2$ values around 1.1. The 1TA and the 1TM give nearly the same spectral parameters.  The cooling rates for the FC and the VC are $19.99_{-10.14}^{+10.14}$ M$_{\odot}$ yr$^{-1}$ and $20.13_{-9.85}^{+10.56}$ M$_{\odot}$ yr$^{-1}$, respectively. Therefore, A2556 could be classified as a weak cool core cluster. This conclusion consists with the cooling time within $1.5^{\prime}$ of A2556, 6.5 Gyr (Gu et al. 2009), according to which A2556 could also be classified as a weak cool core cluster (Mittal et al. 2009).

\section{DISCUSSION}\label{dis}

In \S 2, we present robust evidences for the existence of the weak shock front located at approximately 160$h_{71}^{-1}$ kpc northeast of the cluster center. As in the Perseus, the Hydra A, and several other clusters, the AGN activity might be the cause of the weak shock in A2556. However, partly due to the poor photons statistics, we fail to find robust evidence of AGN bubbles and disturbances on cluster scales. In this section, we shall discuss the other two possible origins of this weak shock front more particularly.

\subsection{ Lack of Supersonic Substructure}
A2556 has a cD galaxy which is consistent with the X-ray peak within $1^{\prime\prime}$ as shown in Figure \ref{ima}b. This indicates A2556 has no violent disturbance caused by major merger in recent $\sim 10^9$ yrs. The difference in photometric magnitudes between two most luminous galaxies, namely the magnitude (luminosity) gap $\Delta mag_{12}$, is the simplest measurable statistic to quantify the dynamical age of a system (Milosavljevi$\acute{c}$ et al. 2006): larger $\Delta mag_{12}$ indicates an older dynamical age.  According to the absolute total magnitudes of the brightest (2MASX J23130142-2138039) and second-brightest (2MASX J 23132498-2137390) galaxies from Smith et al (2004), $\Delta mag_{12}=0.851$ in R band. Comparing it with the mean magnitude gap of 730 clusters (Milosavljevi$\acute{c}$ et al. 2006), $\langle \Delta mag_{12}\rangle \simeq 0.75$, we conclude that A2556 is a dynamical old cluster.

\begin{figure}
    \begin{center}
    \includegraphics[angle=0,scale=0.45]{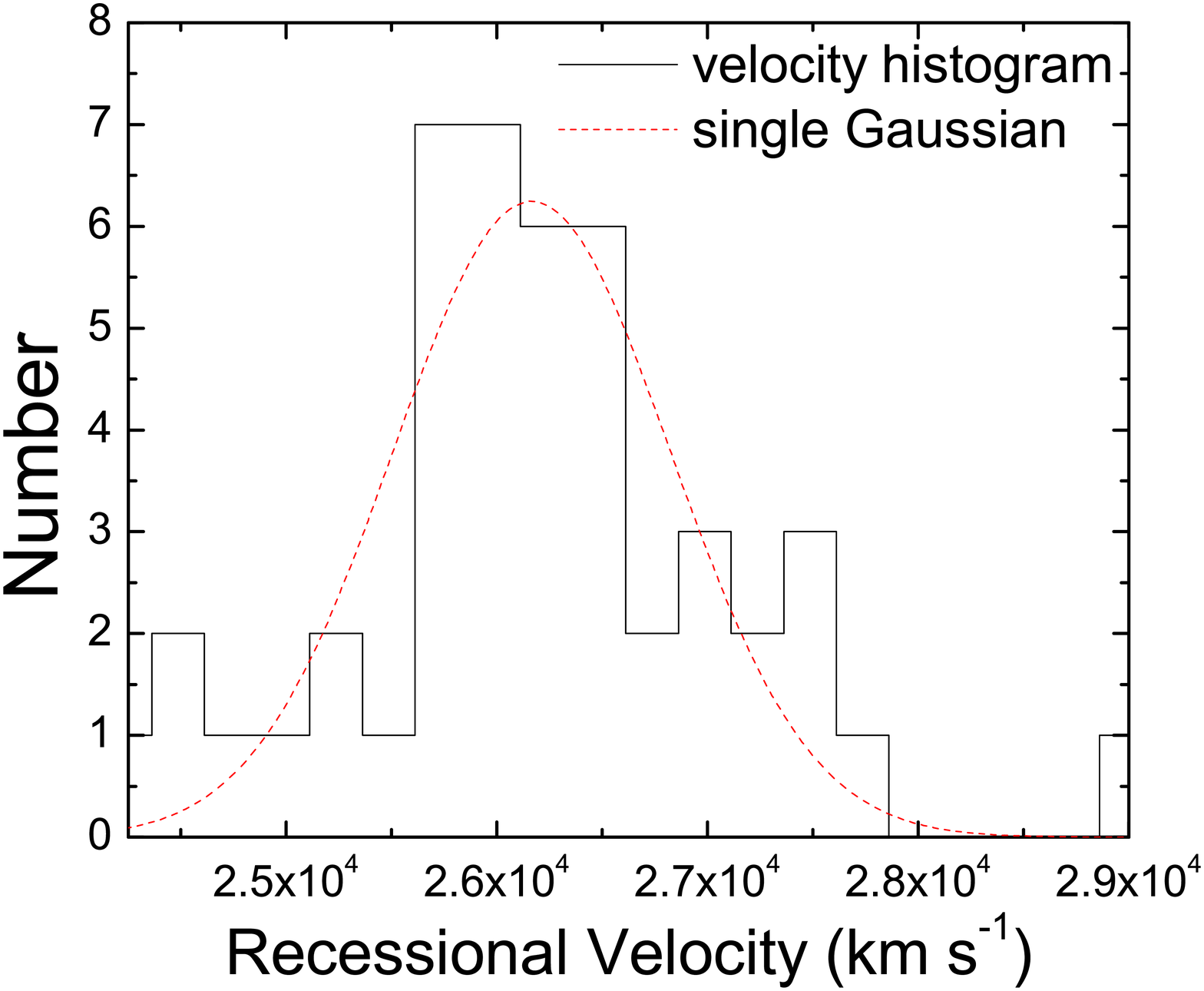}
    \caption{Recessional velocity histogram of 46 member galaxies in A2556, which is featured with a single Gaussian distribution. The best Gaussian fit gives $\chi^2/dof=27.87/17$ and the Kolmogorov-Smirnov statistic tells observed distribution has a probability of 95\% to be a single Gaussian profile, which indicates that A2556 is very unlikely to have subcluster.}\label{dis_v}
    \end{center} 
    \end{figure}

Krick et al. (2007) measured the flux, surface brightness profile, color and substructure in the diffuse intracluster light (ICL) of A2556. They concluded that this cluster is dynamical relaxed and shows no subcluster. We further examine the line-of-sight velocity distribution of 46 member galaxies in A2556 identified by Smith et al. (2004) and plot it in Figure \ref{dis_v}, which shows a single velocity plateau from eye's impression. We can fit the velocity distribution with a single Gaussian profile, the best fit ($\chi^2/dof =27.87/17$) gives the average velocity of $\langle v\rangle = 26165.29\pm88.34$ km s$^{-1}$ and the corresponding variance of $\sigma = 658.31\pm88.37$ km s$^{-1}$. The Kolmogorov-Smirnov statistic tells observed distribution has a probability of 95.0\% to be a single Gaussian profile, which indicates that A2556 is very unlikely to have subcluster.

Previously observed shock fronts in ICM of galaxy clusters, for instance bullet cluster (Markevitch et al. 2002) and A520 (Markevitch et al. 2005), could be attributed to supersonic infalling subclusters. However, there are indications that A2556 possibly has no subcluster. Therefore, the weak shock front in A2556 may not be formed by the supersonic infalling of the subcluster.

\subsection{Nonlinear Steepening Weak Shock in A2556}

Besides the supersonic flow, the nonlinear steepening from normal acoustic wave could also generate shock front (Landau \& Lifshitz 1959). Such nonlinear steepening is a possible reason for the weak shock front in A2556, since the supersonic infalling of the subcluster may not be the origin of the shock front in it.

Firstly,  we would like to discuss the origin of normal acoustic waves in clusters. According to their numerical simulations Nagai et al. (2003) showed that even for the relatively relaxed cluster gas turbulent velocity in the ICM can reach up to about 20\% - 30\% of the local sound speed, and this result is not rule out by current observational upper limits (e.g., Sandsers 2011). Such turbulence generates acoustic wave in the ICM (Fujita et al. 2003), which can be described by Euler's equation. Since the turbulent velocity $v$ is comparable with the sound velocity $c_s$, the nonlinear term $({\bf v}\cdot {\bf grad}){\bf v}$ in Euler's equation cannot be neglected (Landau \& Lifshitz 1959). This leads the acoustic wave to inevitably steepen into the shock wave. According to this mechanism, the ICM turbulence of A2556 could be the origin of the shock front we detected.

Secondly, these shock waves would dissipate the energy and heat the surrounding gas of the clusters, and we perform a schematic calculation for the nonlinear steepening weak shock. Following the method of Stein and Schwartz (1972). When acoustic wave steepens, the crest overtakes the trough in a distance given by
\begin{equation}
\int\frac{\gamma+1}{2} vdt=\int\frac{\gamma+1}{2}\frac{v}{c_s}dL=\frac{\lambda}{2},
\end{equation}
where $\gamma=5/3$ is the adiabatic index for monatomic gas, $\lambda$ is the pulse width, $v$ is the pulse velocity amplitude, $c_s$ is the sound speed (for A2556,  $c_s=900$ km s$^{-1}$ according to its average temperature 3.2 keV), and $L$ is the distance, which the excited sound wave travels before it steepens into the shock front. Since energy flux of acoustic wave $F \propto \rho v^{2}$ and assuming acoustic wave prior to shock formation does not dissipate any energy, we get the relation $v \propto \rho^{-1/2}$. Additionally, with a single $\beta$ distribution $\rho=\rho_{0}[1+(L/r_{c})^{2}]^{-3\beta/2}$ for the gas density in the ICM, where $r_{c}$ is the core radius, we obtain
\begin{equation}\label{lam}
 \lambda = 2\int_{0}^{L}\frac{\gamma+1}{2}\frac{v_{0}}{c_{s}}[1+(L/r_{c})^2]^{3\beta/4}dL,
\end{equation}
where $v_{0}$ is the initial velocity amplitude.
Suppose the acoustic wave originates from the cluster center. The distance it travels before steepens to the shock front is $L \simeq 160h_{71}^{-1}$ kpc in A2556. According to Eq. \ref{lam}, $\lambda \simeq 400h_{71}^{-1}$ kpc, which means the period of acoustic wave generated by the turbulence of ICM in cluster center is $\simeq$ 0.43 Gyr.

\section{SUMMARY}\label{sum}

The \chandra\ observation of A2556 reveals a weak shock front located at approximately 160$h_{71}^{-1}$ kpc northeast of the cluster center with $\mathcal{M}=1.25_{-0.03}^{+0.02}$. Since no subcluster associated with the weak shock front is observed in the optical or X-ray band, we cannot ascribe the formation of the shock front to the supersonic infalling of the subcluster. We suggest the steepening of the acoustic wave could be a possible origin of such shock front, and such acoustic wave could induce by the turbulence of the ICM in the cluster core. 

The nonlinear steepening shock front should be an ubiquity phenomenon in the ICM. In contrast, the observation of nonlinear steepening weak shock front is rare. This may ascribe to that the weak shock wave is difficult to observe. However, to solve this question requires further investigation, including both deeper X-ray observation and extensive theoretic studies.

\section*{Acknowledgments}
We thank Hakon Dahle for kindly providing us with the magnitudes of member galaxies in A2556 that are used in this paper. We thank the anonymous referee for useful suggestions on the manuscript. Also, we thank Chunsheng Pei for helpful suggestions. This work was supported by the Ministry of Science and Technology of the People's Republic of China (973 Program; Grant Nos. 2009CB824900 and 2009CB824904), the National Science Foundation of China (Grant Nos. 10878001, 10973010, and 11125313), and the Shanghai Science and Technology Commission (Program of Shanghai Subject Chief Scientist; Grant Nos. 12XD1406200 and 11DZ2260700).

\begin{deluxetable}{lcccccc}
\centering
\tabletypesize{\scriptsize} 
\tablecaption{Four Spectra Model Fittings for A2556 in \S\ \ref{cla}.} 
\tablewidth{0pt} 
\tablecolumns{10} \tablehead{

    \colhead{Model\tablenotemark{a}} &  \colhead{T} & \colhead{Z} &  \colhead{T$_{min}$} &  \colhead{$\dot M$} & \colhead{$\chi^2/{\rm dof}$}\\
  \colhead{}  & \colhead{(keV)} & \colhead{$Z_\odot$} & \colhead{(keV)} & \colhead{($M_{\odot}$ yr $^{-1}$)} &   \colhead{} }
   \startdata
    1TA  & 3.21$\pm0.09$ & 0.43$^{+0.06}_{-0.05}$ & ${--}$ & ${--}$ & 1.13 \\
    1TM  & 3.25$\pm0.01$ & 0.39$\pm0.05$ & ${--}$ & ${--}$ & 1.13 \\
    FC  & 3.46$\pm0.15$ & 0.44$\pm0.06$ & 0.10 & 19.99$\pm10.14$ & 1.10 \\
    VC  & 3.46$\pm0.15$ & 0.43$\pm0.06$ & 0.15$^{+0.06}_{-0.05}$  & 20.13$^{+10.56}_{-9.85}$ & 1.10 \\

\enddata
\tablenotetext{a}{1TA represents WABS$\times$APEC model; 1TM represents WABS$\times$MEKAL model; For the MEKAL$+$MCKFLOW model, FC represents the lower bound on the gas temperature in the MCKFLOW fixed at 0.1 keV and VC represents this component set free.}
\end{deluxetable}

\clearpage

  \end{document}